\documentclass[sigconf,authorversion]{acmart}
\AtBeginDocument{%
  \providecommand\BibTeX{{%
    \normalfont B\kern-0.5em{\scshape i\kern-0.25em b}\kern-0.8em\TeX}}}

\copyrightyear{2023}
\acmYear{2023}
\setcopyright{acmlicensed}
\acmConference[SC-W '23]{Workshops of The International Conference on High Performance Computing, Network, Storage, and Analysis}{November 12--17, 2023}{Denver, CO, USA}
\acmBooktitle{Proceedings of the SC '23 Workshops of The International Conference on High Performance Computing, Network, Storage, and Analysis (SC-W '23), November 12--17, 2023, Denver, CO, USA}
\acmDOI{10.1145/3624062.3626283}
\acmISBN{979-8-4007-0785-8/23/11}

\begin{document}

\title[The Common Workflow Scheduler Interface: Status Quo and Future Plans]{The Common Workflow Scheduler Interface:\\ Status Quo and Future Plans}

\author{Fabian Lehmann}
\email{fabian.lehmann@informatik.hu-berlin.de}
\orcid{0000-0003-0520-0792}
\affiliation{%
  \institution{Humboldt-Universität zu Berlin}
  \streetaddress{Unter den Linden 6}
  \city{Berlin}
  \country{Germany}
  \postcode{10099}
}
\author{Jonathan Bader}
\email{jonathan.bader@tu-berlin.de}
\orcid{0000-0003-0391-728X}
\affiliation{%
  \institution{Technische Universität Berlin}
  \streetaddress{Straße des 17. Juni 135}
  \city{Berlin}
  \country{Germany}
  \postcode{10623}
}
\author{Lauritz Thamsen}
\email{lauritz.thamsen@glasgow.ac.uk}
\orcid{0000-0003-3755-1503}
\affiliation{%
  \institution{University of Glasgow}
  \streetaddress{University Avenue}
  \city{Glasgow}
  \country{United Kingdom}
  \postcode{G12 8QQ}
}
\author{Ulf Leser}
\email{leser@informatik.hu-berlin.de}
\orcid{0000-0003-2166-9582}
\affiliation{%
  \institution{Humboldt-Universität zu Berlin}
  \streetaddress{Unter den Linden 6}
  \city{Berlin}
  \country{Germany}
  \postcode{10099}
}

\renewcommand{\shortauthors}{Lehmann et al.}

\begin{abstract}
Nowadays, many scientific workflows from different domains, such as Remote Sensing, Astronomy, and Bioinformatics, are executed on large computing infrastructures managed by resource managers.
Scientific workflow management systems (SWMS) support the workflow execution and communicate with the infrastructures' resource managers.
However, the communication between SWMS and resource managers is complicated by a) inconsistent interfaces between SMWS and resource managers and b) the lack of support for workflow dependencies and workflow-specific properties.

To tackle these issues, we developed the Common Workflow Scheduler Interface (CWSI), a simple yet powerful interface to exchange workflow-related information between a SWMS and a resource manager, making the resource manager workflow-aware.
The first prototype implementations show that the CWSI can reduce the makespan already with simple but workflow-aware strategies up to 25\%.
In this paper, we show how existing workflow resource management research can be integrated into the CWSI. 

\end{abstract}


\maketitle
\textit{This work was presented at the 18th Workshop on Workflows in Support of Large-Scale Science (WORKS 2023) and was published as part of the workshop paper ``Novel Approaches Toward Scalable Composable Workflows in Hyper-Heterogeneous Computing Environments'' in the Proceedings of the SC '23 Workshops of The International Conference on High Performance Computing, Network, Storage, and Analysis (SC-W '23) \url{https://doi.org/10.1145/3624062.3626283}}
\section{Introduction}
Analyzing large datasets is the daily business of many scientists~\cite{yates2021reproducible, garcia2020sarek, muirRealCostSequencing2016,rs13061125, Sudmanns_Tiede_Augustin_Lang_2019}.
The data analysis often involves multiple dependent steps, which can be organized as a workflow~\cite{lehmannFORCENextflowScalable2021}.
As these workflows are becoming increasingly complex and datasets easily exceed hundreds of gigabytes or even terabytes~\cite{alam2022challenges,muirRealCostSequencing2016}, scientists use scientific workflow management systems (SWMS), such as Nextflow, Airflow, or Argo, and computer clusters.
One essential feature of SMWSs is communication with a resource manager, such as SLURM, Kubernetes, or OpenPBS.
Therefore, the SWMS submits ready-to-run tasks to the resource manager, and the resource manager takes over the responsibility for assigning these tasks to a node that executes them.
This simplifies the workflow execution on large-scale computing infrastructures and hides the complexity from the scientist.
However, as with SWMS, there is also a variety of resource managers available, and different clusters may use a different one.
In a worst-case scenario, the SWMS preferred by the scientist does not support the cluster's resource manager at all.
Even if the SWMS supports a given resource manager, features beyond submitting tasks and awaiting their completion are frequently not supported.

In this paper, we first give an overview of the Common Workflow Scheduler (CWS) and the Common Workflow Scheduler Interface (CWSI) which we both first presented in ~\cite{lehmannHowWorkflowEngines2023}.
The CWSI is used to exchange workflow-related information between SWMSs and resource managers.
Second, we present prior results, showing promising outcomes when using the CWSI with workflow-aware resource management methods.
Moving on, we outline SWMS, where we started to implement CWSI support and demonstrate how the CWS can serve as a central point for provenance.
Last, we illustrate how the CWS can be extended with new scheduling, resource allocation, and runtime prediction methods.
\section{Common Workflow Scheduler}
In the present landscape, each resource manager has its own unique way of handling task submissions. 
For example, a task's definition significantly differs between SLURM and Kubernetes.
While SLURM supports task dependencies, Kubernetes lacks this feature.
To address the challenge that resource managers schedule workflow tasks without workflow awareness, we developed the Common Workflow Scheduler (CWS)~\cite{lehmannHowWorkflowEngines2023}.
The CWS allows for the transfer of essential information, such as input files, CPU, and memory requests, along with task-specific parameters using the Common Workflow Scheduler Interface (CWSI).
Task-specific parameters vary for each task invocation and are passed on to the utilized tools.
For further details on the interface, we refer to our previous paper~\cite{lehmannHowWorkflowEngines2023}.

\begin{figure}[t!]
  \centering
  \includegraphics[width=\columnwidth,trim={78.0mm 77.0mm 88.0mm 20.0mm},clip]{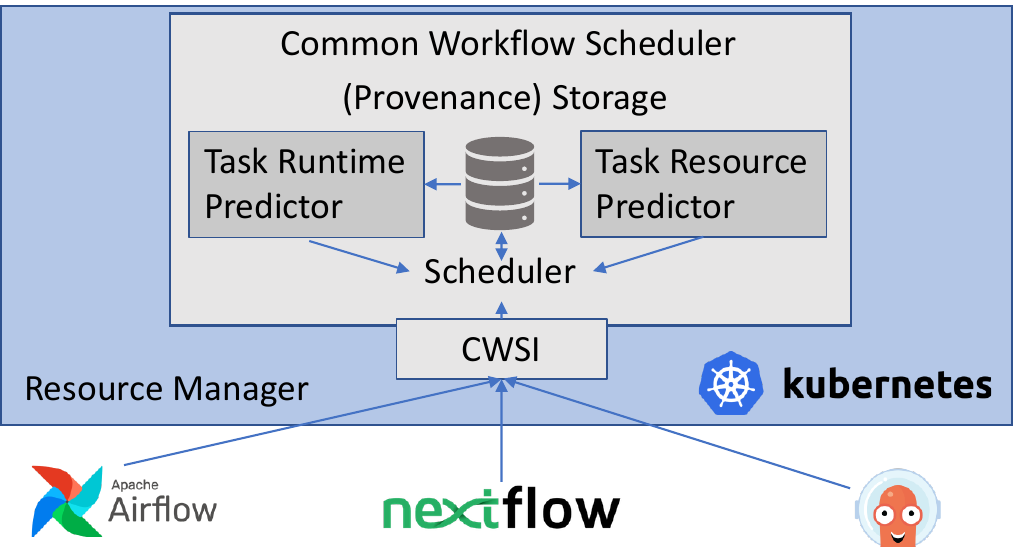}
  \caption{Architecture overview: The Common Workflow Scheduler with its interface and task runtime and task resource predictor component for Kubernetes as an exemplary resource manager.}
  \label{fig:overview}
\end{figure}

In Figure~\ref{fig:overview}, we provide an architectural overview for a single resource manager, in this case, for Kubernetes.
The CWS runs as a component in the resource manager and exposes the CWSI.
A resource manager has to implement the CWS with its interface once.
Conversely, a workflow engine needs to implement support for CWSI to work with all resource managers already offering CWSI.
SWMSs such as Airflow, Nextflow, or Argo send their requests, which are then kept in memory of CWS.
From this storage, the CWS can fetch the workflow graph and task dependencies and use this information for scheduling.
This storage can further be used for provenance to trace the workflow execution; we elaborate on this in Section~\ref{sec::prov}.
The CWS can be extended with task runtime and resource predictors that read task information from the storage and learn characteristics.
Such learned characteristics can then be used to predict the demands for upcoming tasks, which is helpful for better scheduling.
We provide examples for such prediction strategies in Section~\ref{sec::resourceManagement}.
Notably, workflow engines with CWSI support do not need their own scheduler component. 
Instead, all ready-to-run tasks are submitted to the resource manager and the scheduling happens there.

\begin{figure*}[t!]
  \centering
  \includegraphics[width=.962\textwidth,trim={2.5mm 2.8mm 2.5mm 2.5mm},clip]{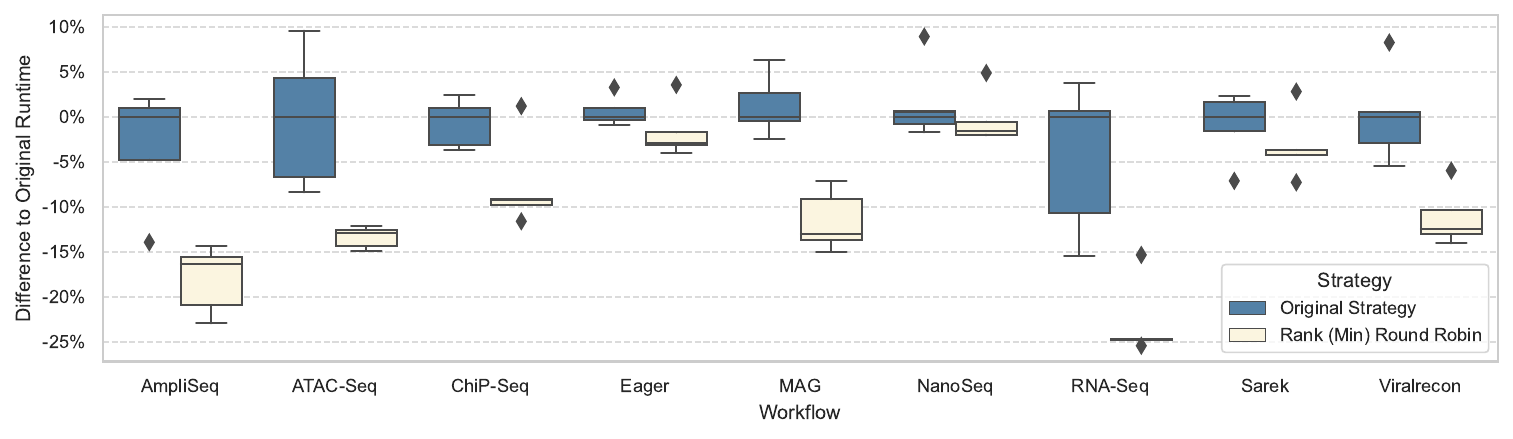}
  \caption{The runtime difference between the original runs' median and the Rank (Min) Round Robin scheduling for the nine most popular nf-core workflows.}
  \label{fig:firstSchedulingResults}
\end{figure*}

We have implemented a plugin\footnote{\url{https://github.com/CommonWorkflowScheduler/nf-cws}} for the SWMS Nextflow to communicate with the CWSI and the CWS for the resource manager Kubernetes\footnote{\url{https://github.com/CommonWorkflowScheduler/KubernetesScheduler}}.
Figure~\ref{fig:firstSchedulingResults} shows the results from running nf-core workflows with the original Nextflow-Kubernetes interaction (Original strategy) and the Rank (Min) Round Robin scheduling algorithm.
nf-core is a collection of best-practice Nextflow workflows which all come with small test sets.
The Rank (Min) Round Robin, on average, outperformed other strategies tested with a median runtime improvement of up to 24.8\% and an average reduction of 10.8\% compared to the original strategy~\cite{lehmannHowWorkflowEngines2023}.

\section{SWMS support}
We started by implementing the CWSI for a resource manager, Kubernetes, and a SWMS, Nextflow. 
We are now actively working to extend our project to support other popular SWMSs, namely Airflow and Argo, to further explore and demonstrate the benefits of the CWSI.
Below, we describe these three SWMSs and discuss the integration of the CWSI.
\\\\
\textit{Nextflow} is a workflow engine initially designed for bioinformatics but getting uptake also in different domains and is used by more than 1,000 organizations~\cite{ditommasoNextflowEnablesReproducible2017,tommasoQuickOverviewNextflow2022,lehmannFORCENextflowScalable2021}.
One of the main advantages of Nextflow is its support for, at the time of writing, 20 different resource managers.
The large support makes it easy to port Nextflow workflows between environments.
The support is achieved by abstracting the resource manager from the scientist but also from the internal Nextflow logic.
Accordingly, Nextflow only supports the basic features of resource managers.
For example, on SLURM, the task dependency feature is not used.
Thus, Nextflow can profit from providing additional workflow context to the resource manager.
\\\\
\textit{Airflow} is an Apache Incubator project designed for workflow management. 
Similar to Nextflow, Airflow supports Kubernetes as a resource manager and is also not exclusively tied to it.
Airflow supports workflow-aware scheduling for Kubernetes through a tailor-made strategy exclusively implemented for the Airflow-Kubernetes interplay.
Therefore, Airflow starts a big worker on every node for the whole workflow execution and assigns tasks into these worker pods bypassing Kubernetes' task assignment logic.
However, this strategy has a significant drawback: the big containers will request resources for the entire workflow execution time regardless of the actual load.
As many workflows have a merge point somewhere, where the entire execution is waiting for one particular task, this strategy leads to substantial resource wastage.
By integrating the CWSI into Airflow, we aim to retain its workflow-aware scheduling capabilities while preventing unnecessary resource requests throughout the runtime. 
This optimization ensures more efficient utilization of resources and minimizes wastage on a large scale.
One big difference to our already existing Nextflow interaction is the knowledge of the physical DAG in Airflow.
While this was foreseen in the development of the CWSI, we have to make use of it in our CWS implementation.
\\\\
\textit{Argo} is a SWMS designed exclusively for Kubernetes. 
However, since Kubernetes lacks support for task dependencies, Argo also submits each task individually, and Kubernetes then schedules them in a FIFO manner.
This is comparable to the strategy of Nextflow and, thus, makes Argo an ideal candidate to support our CWSI.
Just like Nextflow, Argo is expected to benefit in a similar way.
We are currently working on developing an Argo extension to achieve this.
\section{Provenance with the CWSI}\label{sec::prov}
Workflow provenance is one aspect that needs to be addressed in all SWMS~\cite{baderAdvancedMonitoringScientific2022,alam2022challenges,davidson2008Provenance}.
Since the CWSI takes a central role in workflow execution, possessing comprehensive knowledge of the resource manager and the SWMS, it emerges as the most suitable entity for the management of provenance data.

All SWMS represent provenance differently, so it is very heterogeneous~\cite{sergio2009ProvTaxonomy}.
Further, resource managers and SWMSs are only designed to gather a portion of the available data, each focusing on collecting data in its own scope~\cite{baderAdvancedMonitoringScientific2022}.
Accordingly, the resource manager traces the node states while the SWMS collects task-related metrics.
The CWSI is particularly implemented for each resource manager and can support a resource manager's specific APIs to collect traces while it has knowledge about the workflow.
By gathering and storing all metrics and task dependencies in a centralized manner, provenance becomes more streamlined and manageable.

Another significant advantage of using the CWSI for provenance is that the data will be available across different SWMS, even if a particular SWMS does not yet provide built-in provenance data.
This interoperability ensures that provenance information can be maintained consistently and comprehensively, enhancing workflow traceability and reproducibility. 
In turn, researchers and scientists can have greater confidence in the reliability and trustworthiness of their results.
\section{Advanced Resource Management with the CWSI}\label{sec::resourceManagement}
As we saw in the previous section, the CWS provides information about task executions and performance metrics.
Using this information allows possible interface extensions to derive task characteristics from it.
Task characteristics can be predicted runtime, CPU or memory usage, which can be used for scheduling and fed back to the SWMS.
Many scheduling strategies, such as HEFT~\cite{heft}, require knowledge of this.

In the following, we will show how the CWSI can be used to implement approaches for task resource prediction, task runtime prediction, and scheduling with real workflow systems.

\paragraph{Task resource prediction:} 
Predicting the resources a task instance will utilize enables workflow performance optimization by reducing resource wastage and increasing performance~\cite{tovar2017job}.
Several research approaches tackle this challenge by analytic methods, regression models, or reinforcement learning and achieve a significant reduction in resource wastage~\cite{tovar2017job,witt2019feedback,phung2021not, bader2022RL,tovar2022dynamic}.
A key challenge is to avoid underprovisioning of resources, as this leads to task failures while overprovisioning leads to high resource wastage~\cite{tovar2017job}.
These approaches frequently assume a relationship between input data size and a task's resource usage to predict peak memory consumption, i.e., a task's memory usage increases with bigger inputs. 
Further, many of these approaches conduct a form of online learning, incorporating monitoring data from task executions as feedback.

The CWSI provides information to train such models, e.g., the number of file inputs, input sizes, or peak memory, which are retrieved and stored from monitoring.
As these metrics are constantly gathered and updated, also online learning approaches are applicable.
Therefore, we plan to integrate existing task resource prediction methods in our CWSI prototype to a) increase workflow performance and b) evaluate them under real-world conditions.

\paragraph{Task Runtime Prediction:} Predicting task runtimes is essential as many resource management techniques, such as scheduling, rely on accurate runtime estimates beforehand.
To this end, many existing research approaches rely on historical data to build prediction models~\cite{hilman2018task,pham2017predicting,nadeem2017modeling}.
Many of them build on machine-learning models like neural networks, clustering methods, or regression methods~\cite{hilman2018task, pham2017predicting,da2015online,da2013toward,nadeem2017modeling}.
While, especially complex models, showed to achieve low prediction errors, they also require a lot of training data.
As an alternative, we recently presented Lotaru~\cite{bader2023lotaru}, an online approach that can cope with cold-start problems and is able to predict task runtimes without historical traces.
To do this, Lotaru executes microbenchmarks and quickly runs the workflow with reduced input data locally.
Next, it predicts a task's execution time using a Bayesian linear regression based on the data points collected from the local workflow profiling and the microbenchmarks.

Since Lotaru and other research approaches that support heterogeneous infrastructures require machine characteristics, we are extending our CWSI to store such information and extend the prototype to gather these metrics with Kubestone\footnote{\url{kubestone.io}}. 
We are currently incorporating Lotaru into the CWSI prototype to handle unknown workflows or workflows with a lack of historical data.
Further, we plan to implement other research methods that perform better with more training data provided by the provenance store of CWSI.

\paragraph{Workflow Task Scheduling:} Applying sophisticated scheduling algorithms helps to achieve optimization objectives such as a make\-span reduction, cost reduction, or energy reduction.
Although extensive research in this field exists, many approaches are missing uptake in real-world scenarios.
For instance, Yarn schedules tasks in a fair manner~\cite{tang2016fair}, while Kubernetes applies a Round-robin-like strategy~\cite{carrion2022kubernetes}.
Due to the dynamic nature of workflows and infrastructures, in practice, only dynamic scheduling approaches should be considered, i.e., approaches that can adjust their execution plans or react to failures in the infrastructure.
Some of these dynamic approaches~\cite{pheft, yuAdaptiveReschedulingStrategy2007} are based on the static heuristic HEFT~\cite{heft} and require knowledge about task runtimes and communication times between the nodes.
Our own prior work Tarema~\cite{baderTarema2021}, does not require such metrics but dynamically classifies incoming tasks according to their resource usage to select the best-fitting node.

The CWSI, together with task runtime and resource prediction, provides additional information to apply more sophisticated scheduling techniques. 
We are currently implementing the Tarema strategy into our CWSI prototype and plan other more sophisticated approaches enabled through the additional data provided by the CWSI and their plugins.

\section{Conclusion}
In this paper, we presented the status quo of the Common Workflow Scheduler Interface and described the available plugin for Nextflow and the integration into Kubernetes.
Further, we have demonstrated that by implementing the CWSI alongside basic scheduling approaches like rank and file size, we achieve an average runtime reduction of 10.8\%.
Next, we outlined upcoming support for the workflow engines Airflow and Argo and how to extend the storage to become the central place for workflow provenance.
Additionally, we presented our next steps to implement resource allocation, runtime prediction, and new scheduling methods.
We assume that the planned workflow algorithms that consider cluster heterogeneity and task runtime, as we outlined in this paper, will further improve resource efficiency.

\begin{acks}
This work was funded by the German Research Foundation (DFG), CRC 1404: "FONDA: Foundations of Workflows for Large-Scale Scientific Data Analysis."
\end{acks}

\bibliographystyle{ACM-Reference-Format}
\bibliography{sample-base}

\end{document}